\documentclass[reprint,aps,prl,twocolumn,showpacs]{revtex4-1}

\usepackage{amsmath,graphicx,dcolumn,hyperref}
\usepackage{newtxtext,newtxmath}
\hypersetup{colorlinks=true,citecolor=blue,urlcolor=blue,linkcolor=blue}

\newcolumntype{d}[1]{D{.}{.}{#1}}

\let\originalleft\left
\let\originalright\right
\renewcommand{\left}{\mathopen{}\mathclose\bgroup\originalleft}
\renewcommand{\right}{\aftergroup\egroup\originalright}

\usepackage{tikz}
\usetikzlibrary{matrix,decorations.pathreplacing,calc}
\usepackage{mathtools}
\DeclarePairedDelimiter\bra{\langle}{\rvert}
\DeclarePairedDelimiter\ket{\lvert}{\rangle}
\DeclarePairedDelimiterX\braket[2]{\langle}{\rangle}{#1 \delimsize\vert #2}

\usepackage{subfigure}
\usepackage{threeparttable}
\usepackage{tabularx}
\graphicspath{ {images/} }

\begin{document}
\frenchspacing

\title{Density-matrix description of partially coherent spin-orbit wave packets produced in short-laser-pulse photodetachment}
\author{S. M. K. Law}\email{sammklaw@gmail.com}
\author{G. F. Gribakin}\email{g.gribakin@qub.ac.uk}
\affiliation{School of Mathematics and Physics, Queen's University Belfast,
Belfast BT7 1NN, United Kingdom}
\date{\today}

\begin{abstract}

We investigate orbital alignment dynamics within the valence shell of atoms in
coherently excited $j=3/2,1/2$ fine-structure manifolds generated by short-pulse
photodetachment of F$^-$, Cl$^-$ and Br$^-$ anions.
Using Keldysh-type theory, we calculate the density matrix of the residual atoms
generated by few-cycle pulses, whose elements determine the populations and
coherence among the electronic states.
Our calculations demonstrate that the degree of atomic coherence can be represented
by a near universal function of the ratio between the pulse duration
$\tilde{\tau}_p$ and the beat period $\tau_{j'j}$ of the atomic system, which allows
one to characterize the coherence generated in atomic states.

\end{abstract}

\maketitle

The development of femtosecond and attosecond laser pulses has allowed the
possibility to generate and observe wave-packet dynamics within atoms and molecules.
For instance, the interaction of molecular systems with short pulses can initiate coherent rotational wave-packet dynamics, allowing the observation of a high degree of alignment along the field polarization direction in pump-probe experiments \cite{Seideman95,Machholm01,Seideman01,Stapelfeldt03,Litvinyuk03,Itatani05,Faisal07,Artamonov08,Calvert10}. For laser pulses with duration in the few-fs range, vibrational wave packets have also been observed \cite{Feuerstein03,Bryan10},
while in the attosecond time domain, electron dynamics have been probed in the valence shell of neutral molecules through the process of high-harmonic generation \cite{Vacher15,Kraus13,Itatani05,Baker06}. Since the timescale of motion is dictated by the energy splitting of the states involved, the observation of coherent dynamics requires excitation by a pulse of comparatively short duration (i.e., large bandwidth).


It is known that the process of ionization by short pulses can produce hole states and electronic wave packets with long-lived coherences. Of particular interest is the orbital alignment effect characterized by localization of the electron density hole along the polarization axis of the laser pulse \cite{Young06,Rohringer09,Goulielmakis10}. Several strong-field pump-probe experiments have revealed evidence of coherent wave packet dynamics within the valence shell of positive ions initiated by pump-pulse ionization of neutral atoms
\cite{Goulielmakis10,Fleischer11,Argenti10,Rohringer09,Worner11,Fleischer11}.
Additionally, experiments have demonstrated wave packet motion within C, Si and Ge
open-shell neutral atoms generated by strong-field detachment of the respective
negative ions \cite{Hultgren13,Eklund13}. The data were analyzed theoretically in \cite{Law16}, which clearly demonstrated a reduction in electronic coherence when the pump pulse duration is comparable to or exceeds the atomic spin-orbit period.
The periodic variation of coherent electron wave packets through the time delay can
also have a significant effect on the probability of sequential double ionization,
as demonstrated in the yields of singly-charged cations obtained from pump pulse
detachment of Ag$^-$ and Al$^-$ \cite{Greenwood03,vanderhart06}.



A proper treatment of coherence within electronic states can be done by full numerical simulations of the laser interaction with the target and calculation of the density matrix of the residual atomic system. Existing developments include time-dependent $R$-matrix methods \cite{Lysaght09,Moore11,Rey14}, the time-dependent configuration-interaction singles method \cite{Pabst11,Pabst16} and a multichannel theory \cite{Rohringer09}. In particular, the work of \cite{Rohringer09} applied a reduced density matrix formalism to describe Ne$^+$ and Xe$^+$ cationic states produced by tunnel ionization, and examined the hole dynamics and the value of the coherence between the doublet fine-structure $j=3/2,1/2$ sublevels. Additionally, Refs. \cite{Pabst11,Pabst16} showed that the level of coherence between the residual cationic states could be further reduced by interchannel coupling resulting from the action of the Coulomb interaction on the photoelectron.

In this work, we apply a semi-analytical Keldysh-type approach (KTA) to the calculation of the density matrix of the atomic states produced by short-pulse photodetachment of halogen negative ions with $np^6$ ground-state configuration. The original Keldysh theory \cite{Keldysh65} and its variants are commonly used to study strong field interactions with matter
(see e.g. \cite{Gribakin97,Becker02,Milosevic06,Shearer11,Popruzhenko14}). They
work particularly well for modelling electron detachment from negative ions where long-range effects are insignificant.
In this paper we use KTA to calculate the amplitudes of photodetachment by a short laser pulse, leading to a neutral atom (F, Cl, or Br) and an electron with a given momentum in the final state. The residual atomic state spin-orbit manifold $^2P_j$ consists of the $j=3/2$ ground state (with total angular momentum projections
$m= \pm 1/2,\, \pm 3/2$) and the excited state $j=1/2$ ($m=\pm 1/2$).
By computing the relevant elements of the density matrix for the residual atom, we analyze the behavior of the spin-orbit state coherence for pulse durations in the 10--100~fs range. Combined with fine-structure splittings of 404.10, 882.35, and 3685.24~cm$^{-1}$ for F, Cl, and Br, respectively  (corresponding to beat periods between 82.5 and 9.05 fs), the calculations cover a wide range of parameter space.
Our results indicate that the degree of coherence is determined by the ratio of the pulse length and the beat period, which can be used to predict the subsequent temporal evolution of the coherently excited electronic wave packets.


Throughout our analysis we assume that the laser pulse is linearly polarized,
with a sine-squared envelope and vector potential ${\bf A}(t)={\bf A}_0\sin ^2(\omega t/2N)\sin \omega t$, and of total duration $\tau_p=2\pi N/\omega$, where
$\omega$ and $N$ are the frequency and number of optical cycles, respectively.
The corresponding pulse duration at full width at half maximum (FWHM) of the intensity is $\tilde\tau_p = 0.364 \tau_p$.
Within the KTA, the detachment amplitude for electron transition from an
initial state $\Psi_0^{jm}({\bf r},t)=\psi _0^{jm}({\bf r})e^{-E_jt}$ into a final (Volkov) state $\Psi_\textbf{p}({\bf r},t)$ with asymptotic momentum $\textbf{p}$, is written in the form
\begin{equation}\label{eq:ampbraket}
A_\textbf{p}^{jm}=-i \int_0^{\tau_p} \int \Psi_\textbf{p}^*({\bf r},t) V_F(t) \Psi_0^{jm}({\bf r},t) d\textbf{r} \, dt,
\end{equation}
where $\textbf{F}(t)=-d{\bf A}/dt$ is the electric field and
$V_F(t)=\textbf{r}\cdot\textbf{F}(t)$ represents the interaction operator (atomic units are used). The integral over time in Eq. (\ref{eq:ampbraket}) is evaluated using the saddle-point method \cite{Gribakin97,Shearer11}, which replaces it by a sum over a set of complex saddle points $t= t_\mu$ satisfying 
$S_\textbf{p}'(t_\mu )=0$, $\text{Im}\,t_\mu >0$, where
\begin{equation}
S_{\textbf{p}}(t)=\frac{1}{2}\int^t [\textbf{p}+\textbf{A}(t')]^2 dt' - E_j t,
\end{equation}
is the classical action of the electron the field, and with initial bound-state energy $E_j<0$.
This gives
\begin{equation}\label{eq:amplitude}
\begin{split}
A_\textbf{p}^{jm} =-(2\pi)^{3/2} B \sum_{\mu=1}^{2N+2} &\sum_{m_lm_s}
\left(\pm\right)^l C_{l m_l s m_s}^{jm} Y_{lm_l}(\hat {\bf p}_\mu) \chi_{sm_s}\\
& \quad \times \frac{\exp[iS_{\textbf{p}}(t_\mu)]}{\sqrt{-iS_{\textbf{p}}''(t_\mu)}},
\end{split}
\end{equation}
where $B$ is the asymptotic normalization constant of the bound-state wave function \footnote{$\psi _0^{jm}({\bf r})\simeq Br^{-1}\exp(-\kappa _jr)\sum _{m_lm_s}C_{l m_l s m_s}^{jm} Y_{lm_l}(\hat {\bf r}) \chi_{sm_s}$, where $\kappa _j=\sqrt{-2E_j}$, and we use
values of $B$ from Ref.~\cite{Gribakin97} and experimental energies $E_j$ from \cite{Andersen99}.}, the alternating sign $(\pm)=\pm 1$ determines the phase of the successive saddle-point contributions for $l=1$ \cite{Gribakin16_comment},
$C_{l m_l s m_s}^{jm}$ are the Clebsch-Gordan coefficients,
$Y_{l,m_l}$ denote the spherical harmonics (evaluated at the corresponding saddle
points), and $\chi_{s,m_s}$ are orthonormal spin functions. Within the $LS$-coupling scheme, the second sum in Eq.~(\ref{eq:amplitude}) is over all $m_l$ and $m_s$ that satisfy $m_l+m_s=m$.

The elements of the density matrix of the residual atom at the conclusion of the
pulse are evaluated from the transition amplitude (\ref{eq:ampbraket}) by the
formula
\begin{equation}\label{eq:denmat}
\rho_{j' m' j m}= \int A_\textbf{p}^{j'm'\ast}A_\textbf{p}^{jm}
\frac{d^3\textbf{p}}{(2\pi )^3},
\end{equation}
where the integration is performed numerically in spherical coordinates, up to the photoelectron energy of $15\omega $. The diagonal elements $\rho_{jmjm}\equiv \rho^{(j,m)}$ are probabilities of populating different final states $jm$ of the atom ($\rho^{(j,m)}=\rho^{(j,-m)}$ for linear polarization), and $w=\sum _{jm}\rho^{(j,m)}$ is the total photodetachment probability. The complex off-diagonal elements determine the coherence between the relevant states. Coherent superpositions can only be formed between atomic states with the same value of $m$ \cite{Rohringer09,Goulielmakis10}, as confirmed in our calculations where the density matrix elements with $m' \neq m$ are found to be numerically small (with relative magnitudes $<10^{-2}$).

The coherence between the $j=3/2$ and $1/2$ sublevels is determined by the off-diagonal element $\rho_{\frac{3}{2} \frac{1}{2} \frac{1}{2} \frac{1}{2}}$.
The degree of wave-packet coherence can be characterised by the ratio of the magnitude of the off-diagonal element to the geometric mean of the
corresponding diagonal elements \cite{Goulielmakis10},
\begin{equation}\label{eq:coherencedegree}
g=\frac{\bigl|\rho_{\frac{3}{2} \frac{1}{2} \frac{1}{2} \frac{1}{2}}\bigr|}
{\sqrt{\rho^{(\frac{3}{2}, \frac{1}{2})}
\rho^{(\frac{1}{2}, \frac{1}{2})}}}.
\end{equation}
Here $g=1$ corresponds to a pure quantum state such as that produced by a short pulse, and $g=0$ indicates an incoherent classical ensemble obtained for a long pulse.

The elements of the density matrix (\ref{eq:denmat}) allow one to describe the subsequent evolution of the residual atom in terms of the time-dependent density operator,
\begin{equation}
\hat{\rho}_a(t)=\sum_{\substack{j'j m}} \rho_{j'mjm}e^{i(E_j-E_{j'})t} \ket{j'm} \bra{jm},
\label{eq:denoperatetime}
\end{equation}
and consider the expectation value $\langle \hat{O}(t)\rangle=\text{Tr}(\hat{O} \hat{\rho}_a)$ for any observable $\hat{O}$. Coherence between the spin-orbit components results in a periodic variation of the electron density.
The evolution of the laser-generated $np^5$ atomic states can be probed similarly to Refs. \cite{Hultgren13,Eklund13}, by applying a laser pulse which predominantly ionizes electrons with $m_l=0$, and by measuring the signal for polarizations parallel and perpendicular to the polarization of the pump pulse, $S_{||}$ and $S_{\perp}$, respectively.
Considering the ratio $S(t)=(2S_{\perp}-S_{||})/(2S_{\perp}+S_{||})$ \cite{Law16}, we find that the oscillations in the $m_l=0$ electron signal display oscillations with the beat frequency $\omega_b=E_{1/2}-E_{3/2}$,
\begin{equation}\label{eq:signal}
S(t)=\bar{S}+\Delta S \cos(\omega_b t+\beta),
\end{equation}
where
\begin{equation}
\bar{S}=\frac{1}{15}\left(3\tilde \rho^{(\frac{3}{2},\frac{3}{2})}
+5\tilde \rho^{(\frac{1}{2},\frac{1}{2})}+7\tilde \rho^{(\frac{3}{2},\frac{1}{2})} \right),
\end{equation}
is the constant alignment offset, and
\begin{equation}\label{eq:deltaS}
\Delta S = \frac{\sqrt{32}}{15} \bigl| \tilde \rho_{\frac{3}{2}\frac{1}{2}\frac{1}{2}\frac{1}{2}} \bigr|,
\end{equation}
is the amplitude of the beats, determined by the normalized density matrix elements
$\tilde \rho_{j'mjm}\equiv \rho_{j'mjm} /w$, and $\beta$ is an additional phase which may result due to difficulties in determining the zero time delay in experiment \cite{Hultgren13,Eklund13}.
Equation (\ref{eq:deltaS}) shows that the amplitude of the beats is proportional to the magnitude of the off-diagonal element describing the coherence of the atomic system. If we assume that the pump pulse is very short, and only detaches $m_l=0$ electrons (cf.~Ref.~\cite{Law16}), the residual atom will be in a pure state with $g=1$.
The populations are then simply determined by the $LS$-coupling
coefficients,
\begin{equation}
\tilde \rho^{(\frac{3}{2},\frac{3}{2})}=0, \quad
\tilde \rho^{(\frac{3}{2},\frac{1}{2})}=\frac{2}{3}, \quad
\tilde \rho^{(\frac{1}{2},\frac{1}{2})}=\frac{1}{3},
\end{equation}
with the maximum beat contrast $\Delta S/\bar S=8/19\approx 0.42$.

\begin{figure}[t!]
\includegraphics[width=3.5in]{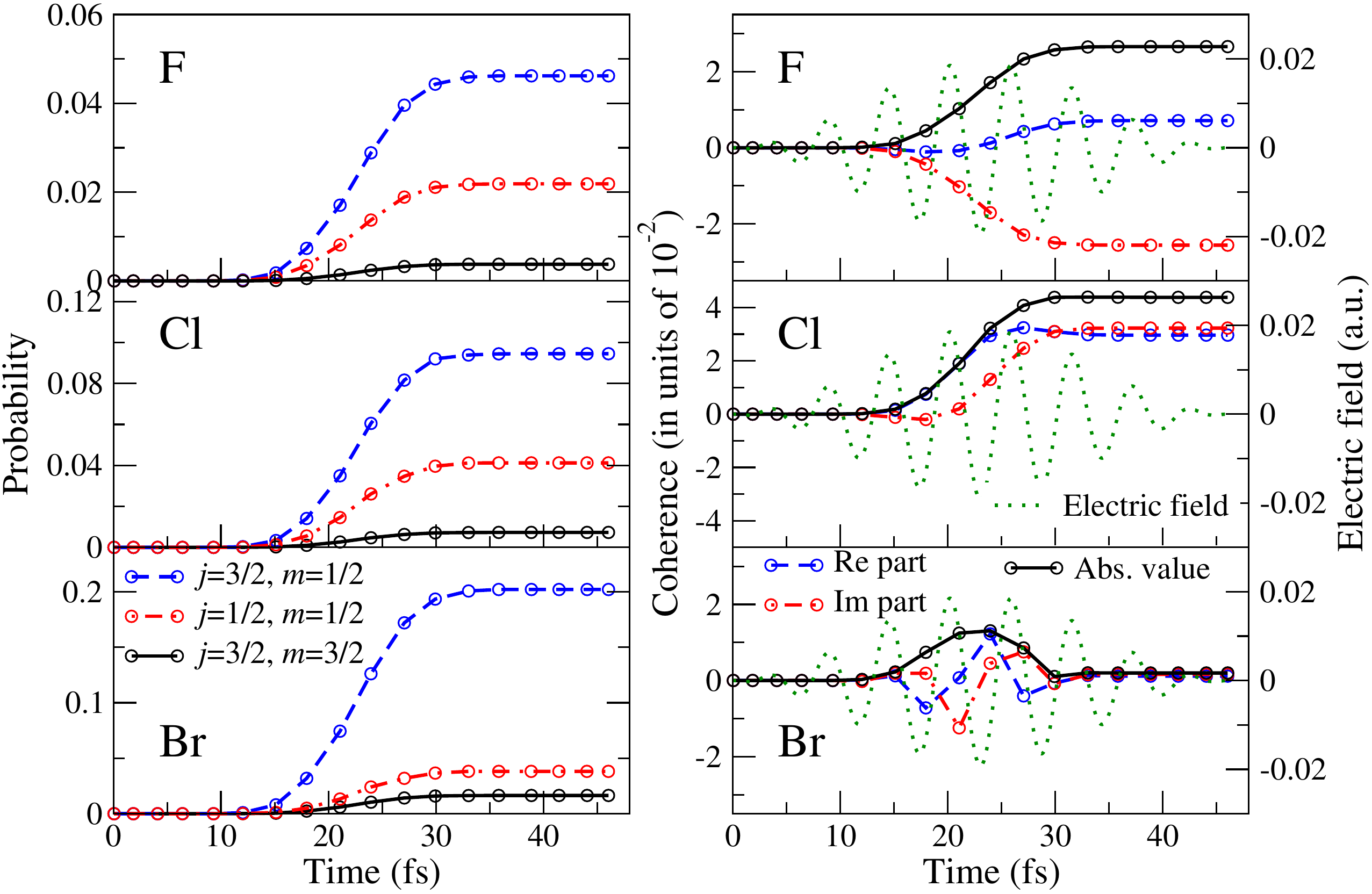}
\caption{Time development of the diagonal (``probability'') and off-diagonal
($\rho_{\frac{3}{2}\frac{1}{2}\frac{1}{2}\frac{1}{2}}$, ``coherence'') elements of the density matrix for F, Cl and Br (top, middle and bottom, respectively) during interaction with an eight-cycle 1800~nm pulse with peak intensity 1.3$\times$10$^{13}$~W/cm$^2$. To guide the eye, values obtained at discrete time instances $\text{Re}\,t'_\mu$ (circles, see text) are connected by the dashed blue line ($\rho ^{(\frac{3}{2},\frac{1}{2})}$ and $\text{Re}\, \rho_{\frac{3}{2}\frac{1}{2}\frac{1}{2}\frac{1}{2}}$), dot-dashed red line ($\rho ^{(\frac{1}{2},\frac{1}{2})}$ and $\text{Im}\, \rho_{\frac{3}{2}\frac{1}{2}\frac{1}{2}\frac{1}{2}}$), and solid black line ($\rho ^{(\frac{3}{2},\frac{3}{2})}$ and $\bigl| \rho_{\frac{3}{2}\frac{1}{2}\frac{1}{2}\frac{1}{2}}\bigr|$), with the electric field of the pulse also superimposed on the right-hand-side panels.}
\label{fig:jpopulation}
\end{figure}

As an example, let us examine photodetachment of F$^-$, Cl$^-$, and Br$^-$ by an eight-cycle pulse with peak intensity $1.3 \times 10^{13}$~W/cm$^2$ and wavelength 1800 nm (the FWHM pulse duration is $\tilde{\tau}_p=17.5$~fs). Figure~\ref{fig:jpopulation} shows the evolution of the diagonal and off-diagonal ($\rho_{\frac{3}{2}\frac{1}{2}\frac{1}{2}\frac{1}{2}}$) density matrix elements. Here the data are presented as a function of the anion-laser interaction time, where the detachment amplitude Eq. (\ref{eq:amplitude}) is computed by discretizing the range of the pulse-anion interaction time $t'$ into $2(N+1)=18$ saddle-point times $\text{Re}\,t'_\mu$ with $0\leq \text{Re}\,t'_\mu \leq \tau_p$, and taking a cumulative partial sum over a subset of saddle-point contributions, $0\leq \text{Re}\,t_\mu  \leq \text{Re}\,t'_\mu $. The discrete time instances $\text{Re}\,t'_\mu$ shown correspond to the electron emission parallel to laser polarization ($\theta=0$), with momentum $p \sim 0.05$~a.u. This allows one to visualize the build-up of the population and coherence during interaction with the field.  

Figure~\ref{fig:jpopulation} shows that the density matrix elements vary rapidly within the central time interval of $\sim $15~fs.
In all cases, the main contribution comes from 6 central saddle points which correspond to three middle cycles of the field. For F, the beat period $\tau _b=2\pi /\omega _b=82.5$~fs is large compared to the pulse length, and the time variation of $\text{Re}\, \rho_{\frac{3}{2}\frac{1}{2}\frac{1}{2}\frac{1}{2}}$ and $\text{Im}\, \rho_{\frac{3}{2}\frac{1}{2}\frac{1}{2}\frac{1}{2}}$ is monotonic. For Cl ($\tau _b=37.8~\text{fs}\approx 2\,\tilde{\tau}_p$), one can detect small nonmonotonic features in both real and imaginary parts of $\rho_{\frac{3}{2}\frac{1}{2}\frac{1}{2}\frac{1}{2}}$. For Br one observes oscillations on the time scale of the spin-orbit period $\tau _b=9.05$~fs, which result in a small final value of the coherence. These oscillations are similar to those seen in the calculations for neon and xenon~\cite{Rohringer09}.

At the end of the 8-cycle, $\tilde{\tau}_p=17.5$~fs, photodetachment pulse, the degree of coherence (\ref{eq:coherencedegree}) is 0.84, 0.70, and 0.02, for F, Cl, and Br, respectively. We see that the value of $g$ becomes progressively smaller for heavier atoms, in particular, producing near incoherent (classical) ensembles for the case of Br. The present calculations clearly demonstrate that large coherence can be observed only for $\tau_b \gtrsim \tilde{\tau}_p$, confirming that higher-bandwidth (i.e., shorter) pulses are required to achieve coherent wave packet formation in heavier systems.

\begin{figure}[t!]
\includegraphics[width=0.4\textwidth]{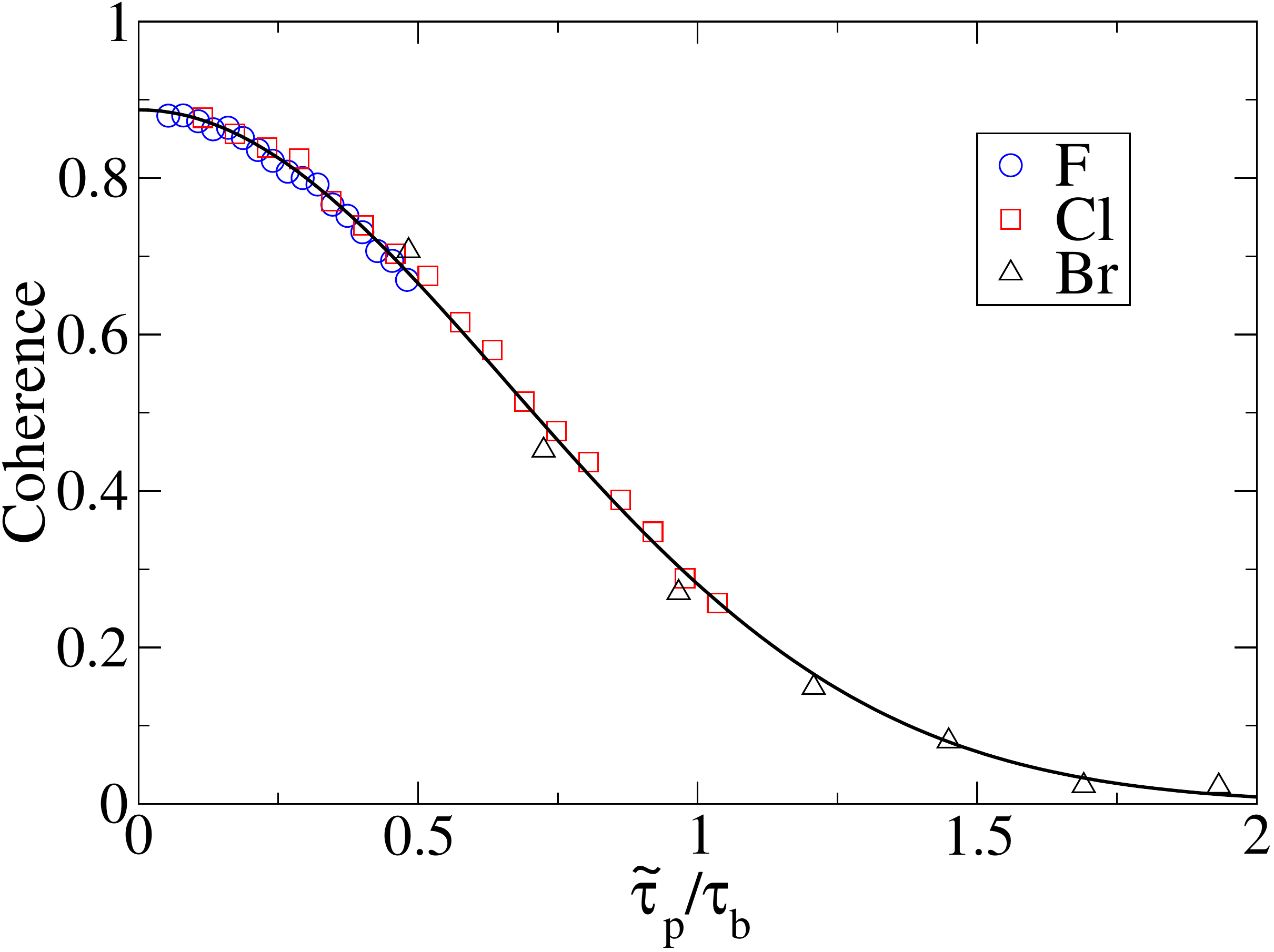}
\caption{The degree of coherence $g$ [Eq. (\ref{eq:coherencedegree})] calculated for
$j=3/2$ and $j=1/2$ states of F (blue circles), Cl (red squares) and Br (black
triangles) produced by short-laser-pulse photodetachment, as a function of the ratio
of the FWHM pulse length $\tilde\tau_p$ to the atomic beat period
$\tau_b$. The laser pulse for each data point consists of an integer number of
cycles $N=2,\,3,\dots ,\,18$ for F and Cl, and $N=2,\,3,\dots,\,8$ for Br, with the carrier wavelength 1800~nm and peak intensity $1.3\times 10^{13}$~W/cm$^2$. The solid line is a Gaussian fit to the data, Eq. (\ref{eq:gaussian}).}
\label{fig:fwhmcoherence}
\end{figure}

To quantify the effect of the pulse duration on the spin-orbit-state coherence of the residual atom, we have performed calculations of the density matrix for a variety of pulse durations $\tau_p$. Using the KTA, we computed values of $g$ for F, Cl, and Br atoms, for $\tau_p=2\pi N/\omega$ with increasing number of cycles $N=2,\,3,\dots $ at fixed frequency $\omega=0.02532$~a.u. (1800~nm wavelength) and peak intensity $1.3 \times 10^{13}$~W/cm$^2$ (the Keldysh parameter in the calculations is within the range $\gamma=0.66$--0.70).
The values of $g$ are presented in Fig. \ref{fig:fwhmcoherence} as a function of the temporal ratio $\tilde\tau_p/\tau_b$ between the FWHM pulse duration and the atomic beat period. The discrete data points correspond to the coherence generated by a laser pulse with $N=2,\,3,\dots,\,18$ cycles for F and Cl, and $N=2,\dots ,\,8$ for Br.

From Fig.~\ref{fig:fwhmcoherence} we see that in spite of the variation in the electron affinity (EA) and large change in the fine-structure splitting (from 1.5\% of EA in F to 13.6\% of EA in Br), the degree of coherence follows a universal dependence on $\tilde\tau_p/\tau_b$ \footnote{A very similar dependence was observed for other intensities ($7.7\times 10^{12}$ and $1.1\times 10^{13}$~W/cm$^2$) and wavelengths (1300~nm), see S. Law, PhD thesis, Queen's University Belfast (2017).}. The solid line in the graph is a curve of best fit to the data, assuming a Gaussian shape,
\begin{equation}\label{eq:gaussian}
g =g_0 \exp\left[-\zeta \left(\tilde\tau_{p}/\tau_b\right)^2\right],
\end{equation}
with two fit parameters, $g_0$ and $\zeta$. The values for the best fit shown in Fig. \ref{fig:fwhmcoherence} are $g_0=0.89$ and $\zeta =1.15$.

The results presented in Fig. \ref{fig:fwhmcoherence} show that for a fixed peak-field strength, the coherence $g$ is a universal function of the scaled pulse length parameter $\tilde{\tau}_p/\tau_{j'j}$, whose shape is close to a Gaussian.
The value of $g$ progressively decreases when the pulse duration $\tilde{\tau}_p$ is
increased (by increasing the number of cycles $N$ or decreasing the wave frequency $\omega$), and drops close to zero when $\tilde{\tau}_p$ exceeds the atomic beat period $\tau_b$. In particular, our calculations predict a reduction
of $g$ to about 0.25 for $\tilde\tau_{p}/\tau_b=1$, while near-complete
incoherence is reached for $\tilde\tau_{p}/\tau_b= 2$. The latter is observable for Br that possesses the shortest beat period $\tau_b=9.05$ fs, which is exceeded by the pulse duration $\tilde\tau_p$ when $N>5$. Additional calculations show that the values of the coherence are only weakly dependent on the laser intensity and wavelength, at least within the tunnelling regime probed by the calculations \cite{Note2}.

It is possible to further illustrate the relation between the coherence
$g$ and the ratio $\tilde{\tau}_p/\tau_b$, by comparing our results with calculations
that employed computationally more demanding methods to other systems. In Refs. \cite{Rohringer09,Goulielmakis10} a time-dependent multichannel theory (TDMT) was used to calculate the density matrix and coherences of the $^2P_j$ ($j=3/2,1/2$) spin-orbit states of the ions Ne$^+$, Kr$^+$, and Xe$^+$, produced in strong-laser-pulse photoionization of the respective noble-gas atoms. The calculations for Ne and Xe \cite{Rohringer09} were for a 4-cycle, constant-amplitude, 800~nm pulse with intensity of $2.1\times 10^{15}$~W/cm$^2$ (Ne) and $2.8\times 10^{14}$~W/cm$^2$ (Xe). For Kr \cite{Goulielmakis10}, 750~nm pulses of FWHM duration of 3.8 and 7.6~fs and peak intensity of $3\times 10^{14}$~W/cm$^2$ were employed.
Table \ref{tab:coherences} displays the degree of coherence $g^\text{TDMT}$ obtained
in \cite{Rohringer09,Goulielmakis10} for Ne$^+$, Xe$^+$ and Kr$^+$, together with
the respective values of the pulse duration $\tilde{\tau}_p$, the spin-orbit period
$\tau_b$ and the ratio $\tilde{\tau}_p/\tau_b$ for each calculation.
The last column (labelled $g^G$) shows estimates of the coherence obtained from the Gaussian fit to our data, Eq. (\ref{eq:gaussian}).

\begin{table}[ht]
\caption{Comparison of the values of the degree of coherence $g$,
Eq.~(\ref{eq:coherencedegree}), between $j=3/2,1/2$ states generated by
strong-field ionization of Ne, Xe and Kr atoms by a laser pulse with duration
$\tilde{\tau}_p$ shown in the second column. The ratio of the pulse duration to the
atomic beat period, $\tilde{\tau}_p/\tau_b$, is given in the third column.
$g^\text{TDMT}$ denotes the coherence values obtained numerically in Ref. \cite{Rohringer09} for Ne$^+$ and Xe$^+$, and Ref. \cite{Goulielmakis10} for Kr$^+$. $g^G$ is the value predicted by Eq.~(\ref{eq:gaussian}) with parameters $g_0=0.89$, $\zeta =1.15$.}
\label{tab:coherences}
\centering
\begin{ruledtabular}
\begin{tabular}{cccccc}
& & & & \multicolumn{2}{ c }{Coherence} \\
\cline{5-6}
Ion & $\tau_b$ (fs) & $\tilde{\tau}_p$ (fs) & $\tilde{\tau}_p/\tau_b$ & $g^\text{TDMT}$ & $g^G$ \\
\hline
 Ne$^+$ & 42.7 & 10.7\footnotemark[1] & 0.25 & 0.82 & 0.83\\
 Xe$^+$ & 3.2 & 10.7\footnotemark[1] & 3.33 & 0.21 & 0.00\\
 Kr$^+$ & 6.2 & 3.8 & 0.61 & 0.60 & 0.58\\
 Kr$^+$ & 6.2 & 7.6 & 1.23 & 0.13 & 0.16\\
\end{tabular}
\footnotetext[1]{Total duration of the constant-amplitude pulse (``rectangular envelope'').}
\end{ruledtabular}
\end{table}

From Table \ref{tab:coherences}, we see that the spin-orbit-state coherences from the TDMT calculations for Ne$^+$ ($\tilde \tau_p/\tau_b=0.25$) \cite{Rohringer09} and Kr$^+$ ($\tilde \tau_p/\tau_b=0.61$ and 1.23) \cite{Goulielmakis10} are in close agreement with our predictions using Eq.~(\ref{eq:gaussian}). This supports the observation in Fig. \ref{fig:fwhmcoherence} that the coherence follows a universal dependence on the scaled pulse duration $\tilde{\tau}_p/\tau_b$, with a shape close to a Gaussian. The value $g^\text{TDMT}=0.21$ for Xe$^+$ ($\tilde \tau_p/\tau_b=3.33$) \cite{Rohringer09} is higher than our prediction $g^G\sim 10^{-6}$. However, the pulse intensity used for Xe in Ref.~\cite{Rohringer09} is such that photoionization saturates well before the end of the pulse, so that effectively, the Xe$^+$ ions are created on a shorter timescale. Indeed, using the value $g=0.21$ in Eq.~(\ref{eq:gaussian}), we find $\tilde \tau _p \approx 1.12 \tau _b = 3.6$~fs, 
by which time, the probability of ionization of Xe is about 80\% (Fig.~2 in Ref. \cite{Rohringer09}). Thus we see that the Gaussian fit, Eq. (\ref{eq:gaussian}), can be used to predict the coherence of spin-orbit states generated by short-laser-pulse photodetachment of anions and ionization of atoms, i.e., for a large range of systems and pulse parameters. 

It is clear from the analysis in Fig. \ref{fig:fwhmcoherence} and Eq.~(\ref{eq:gaussian}) that the beats amplitude $\Delta S$ [Eq. (\ref{eq:deltaS})] is reduced with the increase in the ratio $\tilde{\tau}_p/\tau_b$, viz. 
$\Delta S \simeq (8g/19)\bar S$. (The probability of forming $j,m=3/2,3/2$ states is usually suppressed relatively to that of $m=1/2$ states, because the former requires detachment of $m_l\neq 0$ electrons.) The beats contrast $\Delta S/\bar S$ can also be affected by the difference in the binding energy of the $j=3/2$ and 1/2 levels. Multiphoton detachment rates are sensitive to the binding energy \cite{Gribakin97}, so large fine-structure splittings can lead to a reduction in the population of the more strongly bound $j=1/2$ level, further reducing the beats amplitude (cf. Ref.~\cite{Law16}). Thus, a density matrix calculation of the residual atom using strong-field theory is required in order to fully characterize the electron spin-orbit wave packet.


In conclusion, we have analyzed the coherence in doublet $^2P_j$ spin-orbit wave
packets generated by short-pulse photodetachment of F$^-$, Cl$^-$ and Br$^-$.
By calculating the elements of the density matrix using a Keldysh-type approach, we
examined the dependence of the degree of coherence $g$ on the pulse duration
$\tilde{\tau}_p$ and spin-orbit beat period $\tau_b$. Calculations of the density matrix for a variety of values of $\tau_p$ reveal that the degree of coherence $g$ is a universal function of the ratio $\tilde{\tau}_p/\tau_b$, and exhibits a Gaussian-like decrease with the increase of the pulse duration relative to the timescale of motion. Our analysis confirms that residual atomic states are nearly pure, i.e., are comprised of highly coherent superpositions, for very short pulses, but become close to a classical (incoherent) ensemble whenever $\tilde{\tau}_p/\tau_b>2$.
These findings are in accord with previous experimental \cite{Hultgren13,Eklund13}
and theoretical observations \cite{Rohringer09,Goulielmakis10}.
Our data for the degree of coherence $g$ is also in close agreement with the results
from a numerical multichannel theory \cite{Rohringer09,Goulielmakis10} for the given
values of the ratio $\tilde{\tau}_p/\tau_b$, which provides evidence for the accuracy of the Keldysh approach in modelling the spin-orbit dynamics produced in a short pulse.

The calculation of the density matrix and degree of coherence presented in this
paper can be extended to other systems, e.g., C$^-$, Si$^-$, and Ge$^-$, with $np^3~^4S_0$ ground states and three-level residual atomic states $np^2~^3P_J$ ($J=0,\,1,\,2$). In particular, this will provide a full description of the mixed-state wave packets in the spin-orbit state manifolds of C, Si and Ge produced in femtosecond-pulse photodetachment of the corresponding ions that were studied experimentally \cite{Hultgren13,Eklund13}.
Additionally the KTA could be extended to describe excitation of vibrational wave packets in molecules, induced by short-pulse photodetachment. Although the nuclear dynamics is commonly described using the approximation of instantaneous detachment, the pulse duration in real experiments is often comparable to the characteristic vibrational periods, which calls for a proper density-matrix description of the mixed vibrational states of the residual molecular system.

We thank Daniel Clarke for useful discussions. The work of S.M.K.L. was supported by the Department for Employment and Learning, Northern Ireland.

\end{document}